\documentclass[reprint,eqsecnum,floats,aps,amsmath,amssymb,nofootinbib,onecolumn,doublespacing,showpacs]{revtex4-2}

\usepackage{graphicx}
\usepackage{amsmath,amssymb}
\usepackage{hyperref}
\usepackage{graphicx}
\usepackage{subfigure}
\usepackage{arydshln}
\usepackage{inputenc}
\usepackage{graphicx}
\usepackage{amsmath,amssymb}
\usepackage{hyperref}
\usepackage{appendix}
\usepackage{color}

\begin{document}

\title{Choice of vacuum state and the relation between inflationary and Planck scales}

\author{Maciej Kowalczyk}
\email{maciej.kowalczyk@uwr.edu.pl}
\affiliation{Institute for Theoretical Physics, Faculty of Physics and Astronomy, University of Wrocław, pl. M. Borna 9, 50-204 Wrocław, Poland,}
\affiliation{Instituto de Estructura de la Materia, IEM-CSIC, Serrano 121, 28006 Madrid, Spain}

\author{Guillermo A. Mena Marug\'an}
\email{mena@iem.cfmac.csic.es}
\affiliation{Instituto de Estructura de la Materia, IEM-CSIC, Serrano 121, 28006 Madrid, Spain}

\begin{abstract}
Recent observations about the cosmic microwave background evidence a clear discrepancy between the scale of inflation and the Planck scale expected in the conventional inflationary picture, based on simple chaotic inflationary models. This paper explores a possible resolution of this conflict by considering a slight modification of the standard general relativity scenario, which naturally incorporates a cutoff scale consistent with the observations. This last scale of power suppression appears by the combined effect of the preinflationary background dynamics and an associated initial vacuum state for the cosmological perturbations which differs from the conventional Bunch-Davies state of slow-roll inflation. We revise and correct previous discussions in the literature about the relation between this scale and the scale corresponding to the onset of inflation. The mechanism that produces this cutoff is intimately related to the criterion determining a privileged vacuum state for the primordial perturbations. We adhere to a recent proposal based on an asymptotic diagonalization of the Hamiltonian of the perturbations and leading to a non-oscillatory primordial power spectrum. With this choice, we are able to investigate analytically the model, studying in detail some physically interesting cases. These analytic calculations provide further support to our arguments about the relation between the different scales of the system.   
\end{abstract}

\maketitle

\section{Introduction}

The inflationary paradigm \cite{Baumann}, introduced half a century ago \cite{Guth,Linde,Sta,Sato,Kaza}, has become a cornerstone of modern cosmology. Inflation can explain the observed homogeneity, isotropy, and flatness of our universe and the primordial fluctuations \cite{Mukhanov} needed to produce the anisotropies of the cosmic microwave background (CMB) as well as the large scale structures \cite{Liddle}. In the standard inflationary framework provided by chaotic inflation \cite{Linde2}, the single inflaton field that drives the expansion should have an initial energy density of Planck order. On the other hand, observations of the CMB fit remarkably well with the predictions of a slow-roll phase \cite{Baumann,Liddle} and seem to favor a field with an approximately constant potential during inflation \cite{Planck14,Ijjas}. However, an estimation of the energy density corresponding to this phase indicates a value which is twelve orders smaller than the Planck density (see e.g. Ref. \cite{Liu}). 

This discrepancy has led to suggest a scenario with delayed inflation, in which the slow-roll epoch begins after a transition from a Planck-density dominated regime, with a connecting period between the two scales which is far from slow-roll conditions. Moreover, it has been argued that this delay may produce power suppression at large angular scales. This suppression is compatible with observations pointing towards a lack of angular correlations for angles greater than $60^{\rm o}$ \cite{Hinshaw,Planck,Planck2} and could be described in terms of a (nearly hard) cutoff in the primordial power spectrum (PPS) \cite{Melia,Melia2,Sanchis}. Actually, since the sector of large angular scales is seriously affected by the problem of cosmic variance \cite{Mukhanov}, this lack of power might be just a statistical question. Nonetheless, many works have lately explored more exotic possibilities, for instance, more involved inflationary models, with departures from the slow roll at the onset of inflation. This may happen in scenarios with non-minimal coupling of the inflaton \cite{Tiwari}), with multiple scalar fields \cite{Braglia}, or with unconventional inflaton potentials \cite{Hunt,Qureshi,Raga}. Other works have suggested a primordial origin within general relativity \cite{Contaldi,Cline,Wang,Destri,Destri2,Boya,Jain,Scardi,NFS,Dudas,Kouwn,Chen,Hergt}, by including a preinflationary era from which information may have been transmitted (at least partially) to and beyond the slow-roll period. 

Alternately, other proposals have explored more genuinely quantum explanations to this difference between the inflationary and the Planck scales, relating frequently this issue to a lack of power. This is the case, e.g., of quantum gravity effects \cite{Kiefer,cough,AAN,tango,Chatwin}, a thermodynamic effective behavior of spacetime \cite{Alonso}, or a quantum entanglement of inflationary horizons \cite{Hogan}. In particular, especial attention has been paid to models inspired by loop quantum gravity, a nonperturbative quantization of general relativity with important consequences in cosmology \cite{AS}. Loop quantum cosmology (LQC) extends the inflationary universe to preinflationary eras by replacing the {\sl Big Bang} with a quantum region with Planck physics where the cosmological contraction/expansion bounces \cite{AS,APS}. The effects of LQC in the CMB have been analyzed in recent years \cite{dressed,Morris,hyb,Jorma,tango,EM,Antonio} as an alternative to conventional inflation. 

In principle, extended models of conventional inflation which include a preinflationary phase and a transition to delayed inflation might be a way to match Planck scales with the observed scale of inflation. However, it has been claimed \cite{Liu} that they find problems to explain homogeneity and sustain a natural choice of vacuum state that sets the initial conditions for the cosmological perturbations. The main aim of this article is to show that, if one abandons the assumption that the primordial fluctuations be in a Bunch-Davies state \cite{BunchDavies} during the standard inflationary period, one can actually solve such problems and construct a cosmological model that naturally contains the two aforementioned scales without theoretical tensions. 

An important point in this respect is the correct relation between the scale at the beginning of the slow roll and the cutoff scale in delayed inflation. We will prove that their identification is not generally well founded. The actual relation depends on the choice of vacuum, or equivalently on the initial conditions on the perturbations. Only modes that cross exclusively once the Hubble horizon during the entire preinflationary and inflationary evolution, and do this in the slow roll, can be assured to be well described by a Bunch-Davies-like state at the end of inflation if they started in this state. In fact, this condition determines a scale which is different from that associated to the onset of (delayed) inflation. Therefore, a vacuum  state which is optimally adapted to the dynamics of the cosmological background in all the preinflationary stages can display power suppression at scales where slow-roll inflation has already started. 

A nice example of this fact was discussed in Ref. \cite{EM}, comparing the situation for a kinetically dominated preinflationary period in general relativity (see Ref. \cite{Contaldi}) with a quantum gravity scenario. The choice of a vacuum was motivated by a natural criterion to select positive frequency solutions in the asymptotic ultraviolet sector \cite{BeaDiagonal}. This criterion is based on an asymptotic diagonalization of the Hamiltonian of the perturbations in that sector. The criterion generalizes the choice of Bunch-Davies state to backgrounds that are not quasi de Sitter, and leads to primordial spectra free of superimposed highly oscillating contributions that would pump up power when bin-averaged \cite{EM,BeaPrado}. 

The rest of the paper is organized as follows. We provide a short summary of the conventional inflationary paradigm and the slow-roll approximation in Sec. \ref{Sec2}. In Sec. \ref{Sec3}, we delve into the conceptual framework of delayed inflation. We also discuss some consequences of the presence of two different scales, corresponding to the beginning of the slow roll and the cutoff. In Sec. \ref{Sec4} we present a specific proposal for the choice of vacuum state displaying in practice a cutoff originated at Planck scale, and develop an analytic derivation of the PPS in some examples of physical interest after certain simplification of the background-dependent mass of the scalar perturbations. The results of these analytic investigations support that our arguments are correct. Finally, we conclude in Sec. \ref{Sec6} with a summary of our results and a comparison between the analyzed scenario and the particular case of LQC, as an exemplification of the discussed physics. In the following, we adopt by default natural units, setting Newton's constant $G$ and the speed of light $c$ (and Planck's constant $\hbar$ if quantum arguments are involved) equal to one. However, in some occasions, we will employ other systems of units that give a better intuition of the physics discussed, stating this explicitly.  

\section{Cosmological perturbations in de Sitter inflation and slow roll}
\label{Sec2}

In conventional inflationary cosmology, one considers a Friedmann–Lemaître–Robertson–Walker spacetime in general relativity with a matter content given by a single homogeneous scalar field, namely the inflaton. The Friedmann equation and the continuity equation of the matter content govern the evolution of this cosmological background. In conformal time $\tau$ and calling $a$ the metric scale factor, these equations are equivalent to requiring that
\begin{align}\label{Friedman-Equations}
\left( \frac{a^\prime}{a}\right)^2  = \frac{8 \pi }{3}a^2 \rho, \qquad & \qquad \frac{a^{\prime\prime}}{a} = \frac{4 \pi }{3}a^2( \rho  - 3 P),
\end{align}
where the prime denotes the derivative with respect to the conformal time. Consistency of these equations is guaranteed by the condition $\rho^{\prime}=-3a^{\prime}(\rho+P)/a$ on the energy density $\rho$ and the pressure $P$ of the scalar field $\phi$. The explicit expression of these quantities is
\begin{align}
\rho=\frac{1}{2}\left(\frac{\phi^{\prime}}{a}\right)^2+V(\phi), \qquad & \qquad P=\frac{1}{2}\left(\frac{\phi^{\prime}}{a}\right)^2-V(\phi) ,
\end{align}
where $V(\phi)$ is the inflaton potential. Then, the continuity condition on $\rho$ and $P$ implies 
\begin{equation}
\phi^{\prime\prime}+2\frac{a^{\prime}}{a}\phi^{\prime}+a^2\partial_{\phi} V(\phi)=0 .
\end{equation}

To keep our arguments as general as possible, in principle, we will not specify the inflaton potential. During the inflationary phase, we will only admit that it remains approximately constant and dominates over the kinetic contribution to the energy density of the inflaton. This condition is central to the slow-roll approximation. The influence of the potential can be characterized then by two key parameters, $\epsilon$ and $\eta$, which respectively quantify the slope and curvature of this potential. Their definition is
\begin{equation}
\epsilon = \frac{1}{16\pi } \left( \frac{\partial_{\phi} V(\phi)}{V(\phi)} \right)^2, \qquad \eta = \frac{1}{8\pi } \frac{\partial^2_{\phi} V(\phi)}{V(\phi)} .
\end{equation}
For the slow-roll approximation to be good and accurate, these parameters must satisfy the conditions $\epsilon \ll 1$ and $\eta \ll 1$. 

In this scenario, the background scale factor can be well approximated as the solution to the evolution equations \eqref{Friedman-Equations} in the presence of only a cosmological constant, equal to the (nearly constant) value of the potential. The analytic form of the solution is 
\begin{equation}\label{aDesiter}
a(\tau)=a_{inf}\left[1-H_{inf}a_{inf} (\tau-\tau_{inf})\right]^{-1},
\end{equation}
where $a_{inf}$ is the value of the scale factor at the beginning of the inflationary slow-roll phase $\tau_{inf}$, and $H_{inf}$ is the corresponding value of the Hubble parameter. The expression of this parameter is the same as in a de Sitter universe with a cosmological constant given by $V(\phi_{inf})$, according to our comments above, where $\phi_{inf}$ is the value of the inflaton when the slow-roll epoch starts. From the Friedmann equation, we get $H_{inf}^2=8\pi a_{inf}^2 V(\phi_{inf})/3$.

Let us now consider cosmological perturbations on this background, with an eye on their relation with the CMB \cite{Langlois}. The dynamics of the gauge-invariant scalar perturbations are governed by the so-called Mukhanov-Sasaki equation \cite{Mukhanov2,Sasaki,Sasaki2}. Expressing the perturbations in terms of Fourier modes $v_k$ with wavenorm (i.e. angular wavenumber) $k$, this equation can be written in the form
\begin{equation}\label{MSequation}
v_k^{\prime\prime}+(k^2+s)v_k=0,
\end{equation}
where $s$ plays the role of a background-dependent mass. Treating the background during the inflationary epoch as a de Sitter spacetime, in a leading order approximation in which even the influence of the slow-roll parameters is neglected, this mass turns out to be given by $s=- 16 \pi a^2 V(\phi_{inf})/3$ (see e.g. \cite{EM}). Then, employing our expression of the scale factor, we obtain
\begin{equation}
s=- 2 H_{inf}^2 a^2 = - 2  H_{inf}^2 a_{inf}^2  \left[1- H_{inf}a_{inf} (\tau - \tau_{inf})\right]^{-2} .
\end{equation}
With this mass, the general solution to Eq. \eqref{MSequation} is known. Introducing the Fourier scale $k_{inf}= H_{inf}a_{inf}$ and calling $\sigma_{inf}=\tau_{inf}+1/k_{inf}$, we get \cite{Mukhanov}
\begin{equation}\label{solu1}
v_k=A_k \frac{e^{i k(\tau - \sigma_{inf})}}{\sqrt{2k}}\left[1 + \frac{i}{k(\tau  - \sigma_{inf})}\right] + B_k \frac{e^{-i k(\tau  - \sigma_{inf})}}{
\sqrt{2k}}\left[1 - \frac{i}{k(\tau - \sigma_{inf})}\right] .
\end{equation}
The Mukhanov-Sasaki modes are normalized with respect to the Klein-Gordon inner product, so that the corresponding gauge-invariant satisfies canonical commutation relations. This is an important ingredient in the process of constructing a vacuum state for the perturbations. This normalization amounts to the condition \cite{EM}
\begin{equation}
v_k(v_{k}^{\prime})^{*}- v_{k}^{\prime}v_k^{*}=i ,
\end{equation}
where the symbol $^*$ stands for complex conjugation. Equivalently, the normalization requires that the constants appearing in the general solution \eqref{solu1} satisfy that $|B_k|^2-|A_k|^2=1$. 

The PPS can be obtained from the evaluation of the perturbations at the end of inflation (at time $\tau_{end}$) by the relation \cite{Langlois}
\begin{equation}
\mathcal{P}_{s}(k)=\frac{k^3}{2\pi^2}\frac{|v_k(\tau_{end})|^2}{a^2_{end}} ,
\end{equation}
with $a_{end}=a(\tau_{end})$. For a de Sitter background, this PPS can be computed as 
\begin{equation}
\mathcal{P}_{s}(k)=\frac{k^3}{2\pi^2}\lim_{a\rightarrow \infty}\frac{|v_k|^2}{a^2} ,
\end{equation}
an approximation that is good if the number of e-folds during the considered inflationary period is large enough \cite{EM}. Actually, compatibility with observations demands that this is the case, with more than several tens of e-folds \cite{efolds}. This is satisfied even in scenarios with short-lived inflation \cite{Ramirez,Ramirez2,Scacco}. Employing the solution in Eq. \eqref{solu1}, the PPS can then be expressed in terms of the coefficients $A_k$ and $B_k$ as 
\begin{equation}\label{pps}
\mathcal{P}_{s}(k)=\frac{H_{inf}^2}{4\pi^2}|B_k-A_k|^2 .
\end{equation}

The relative phase between $B_k$ and $A_k$ often introduces a rapid oscillatory behavior in $k$ that, in average, leads to an enhancement of power. This is the case when the variation of the phases of these constants is much faster than the variation of their norms \cite{EM}. On the other hand, we notice that different choices of vacuum state for the perturbations are in one-to-one correspondence with different choices of $B_k$ and $A_k$, only restricted by the normalization condition $|B_k|^2=1+|A_k|^2$. As we will discuss in more detail in the next section, it is then possible to select vacua that remove the spurious superimposed oscillations. This is achieved by a Bogoliubov transformation that sends $B_k$ and $A_k$ to their norms, up to an irrelevant common phase. The corresponding PPS becomes
\begin{equation}\label{nopps}
\Tilde{\mathcal{P}}_{s}(k)=\frac{H_{inf}^2}{4\pi^2}\left(|B_k|-|A_k|\right)^2 .
\end{equation}
From the above discussion, it is evident that choosing the correct vacuum state is crucial, as the PPS highly depends on it. In the standard approach, one typically identifies as a natural vacuum the Bunch-Davies state. This state is selected in genuine (cosmological) de Sitter backgrounds by its invariance under the background isometries and its good local behavior \cite{BunchDavies}. By extension, the state is accepted as a privileged vacuum in quasi de Sitter scenarios. This state is obtained with the choice $B_k=1$ and $A_k=0$, and it is straightforward to see from Eq. \eqref{pps} that it leads to a scale invariant PPS.

\section{Delayed inflation and solution to the scale mismatch}
\label{Sec3}

To discuss the compatibility of theoretical predictions with observations, we assume that the framework provided by general relativity is valid up to energy-density scales close to the Planck one, $\rho_P$. We recall, for convenience of the reader, that in MKS units $ \rho_P\approx5.2\times 10^{96} \text{kg m$^{-3}$} $. In a conventional inflationary scenario, within chaotic inflation, this energy density corresponds to a Hubble parameter $H_P$ of Planck order. Again, in MKS units this means $H_P\approx 5.4 \times 10^{43}\textbf{s}^{-1}$.

On the other hand, using the relations $\mathcal{R}=16\epsilon$ and $n_s-1=2\eta-6\epsilon$ between the slow-roll parameter $\epsilon$ and the tensor to scalar ratio $\mathcal{R}$ or the spectral index of the scalar perturbations $n_s$, respectively, we can approximate the PPS by a power law,
writing
\begin{equation}
\mathcal{P}_s(k)=A_s \left(\frac{k}{k_{\star}}\right)^{n_s-1} .
\end{equation} 
Here, $k_{\star}$ is the so-called pivot mode (its value is usually given in inverse megaparsecs, $k_{\star}=0.05\text{Mpc}^{-1}$), and $A_s$ is its amplitude. This PPS determines a Hubble parameter by the formula $ \mathcal{P}_s (k)  \approx {H^2}/{\mathcal{R}}$
\cite{Linde}. Using measurements of the Planck collaboration \cite{Planck}, it is then possible to estimate the Hubble parameter in the slow-roll inflationary phase \cite{Liu}, 
\begin{equation}\label{Hintvalue}
H_{inf}\approx 0.57 \times 10^{-6} H_P 
\end{equation}
(or $H_{inf} \approx 3.1 \times 10^{37}\textbf{s}^{-1}$ in MKS units). This difference of 6 orders of magnitude proves that the inflaton potential needs to decrease dynamically from the end of the Planck regime until the beginning of the slow roll, requiring time for this transition. During it, typically, the universe expands (although not in an exponential way), from a scale factor $a^{(P)}_{end}$ to $a_{inf}$. We will further comment on this expansion later in this section. Note that a contraction in this transition period would impose even more demanding conditions on the inflationary period to reach the degree of homogeneity that is required observationally. 

As we have already commented, the observation of the temperature anisotropies of the CMB has led to accumulated indications of a possible lack of power in the spectrum at angular scales \cite{Hinshaw,Planck,Planck2}. In this respect, it has been proposed that this power suppression might be explained by a (nearly hard) cutoff \cite{Melia,Melia2,Sanchis} at Fourier scales below 
\begin{equation}
k_{min}\approx 6.3 \times 10^{-3} k_{\star} , 
\end{equation}
in terms of the pivot scale (more precisely, we have $k_{min}=(3.14\pm 0.36) \times 10^{-4}\text{Mpc}^{-1}$, in inverse megaparsecs). If this proposed cutoff $k_{min}$ were identified with the smallest wavenorm that crosses the Hubble horizon in the slow-roll epoch, crossing which would occur at the beginning of this inflationary period, we would have that $k_{min}=k_{inf}$ with $k_{inf}= H_{inf}a_{inf}$. For the sake of clarity, let us point out that we place the horizon crossing at the instant when the physical angular wavelength becomes equal to the inverse of the Hubble radius. It was then noted in Ref. \cite{Liu} that the conventional inflationary picture might have troubles to solve the horizon problem. On the one hand, within a standard $\Lambda$CDM (i.e., assuming a matter content after decoupling given by dust matter, radiation, and a cosmological constant), a calculation using the cosmological parameters derived from Planck observations leads to a radial comoving distance of approximately $690/k_{\star}$ (i.e. approximately $1.38 \times 10^4  \text{Mpc}$) for photons travelling to us from the decoupling of the CMB \cite{Liu2}. To avoid horizon problems, the corresponding comoving distance from the Planck regime to decoupling should be at least twice that value. This latter distance is typically dominated by the contribution of the slow-roll inflationary period. In our de Sitter approximation to describe the cosmological evolution in that period, it is easy to conclude that the radial comoving distance for the slow-roll interval is $1/k_{inf}$. Hence, we would need that $k_{inf}\leq 1.45 \times 10^{-3}  k_{\star}$. If we now insist in making $k_{min}$ coincide with the scale at the beginning of the slow roll, we would get that $k_{min}\leq 1.45 \times 10^{-3}   k_{\star}$, which is not fulfilled by approximately {\emph{one order of magnitude}}.

To investigate possible solutions to this conumdrum, the authors of Ref. \cite{Liu} considered a transition from the Planck regime to the inflationary slow-roll era given by an inverse power law in the scale factor, namely $H=B a^{\beta-1}$, where $\beta\leq 1$ and $B>0$ are constants. The restriction to $B>0$ corresponds to expanding behaviors during the aforementioned transition. A cosmological contraction during this epoch would only exacerbate the encountered problems and might even pose a challenge for physical questions such as the thermal evolution of the universe. Besides, the possibility of a preliminary epoch of exponential expansion during the Planck regime was also contemplated, so that the scale factor at the end of this regime, $a^{(P)}_{end}$, could be greater than something of Planck order. In spite of this, the problems persist if the cutoff scale is identified with the scale at the beginning of slow-roll inflation \cite{Liu}. 
 
Therefore, it should be clear that a possible way out of this conflict consists in avoiding this identification. Actually, the (nearly hard) cutoff must correspond to the minimum wavelength in which the perturbations display the same behavior as if they were in a Bunch-Davies state during the slow roll. It is at this point where the relation between the choice of vacuum state and the value of the cutoff emerges. General arguments strongly support that the perturbations have a Bunch-Davies behavior for modes which only experience one horizon crossing during the evolution and this crossing occurs in the slow-roll regime. However, for modes such that this description fails, one should not expect that the Bunch-Davies state is the appropriate vacuum. For modes which experience more than one horizon crossing, there will be excitations when they re-enter the horizon, so that the perturbations may cease to adjust to the Bunch-Davies solution. Moreover, for modes that only exited the Hubble horizon once, but did so during the Planck regime, a description in terms of the Bunch-Davies state adapted to the slow-roll evolution may also be invalid. The specific behavior of the modes in this interval of Fourier scales will depend on our criterion for the determination of the most natural state to play the role of a vacuum \cite{Contaldi,Morris,EM}. 

The above arguments suggest that the minimum scale for which a Bunch-Davies behavior should be retained during the slow roll corresponds in fact to a different scale than the one associated with the beginning of the conventional inflationary period. An optimal possibility, in agreement with the discussion of the above paragraph, is that the cutoff is at the lower end of the sector compatible with the scale invariance of the spectrum, proper of the Bunch-Davies state. This is simply the scale at the end of the Planck regime, given by $k_0= H_{P}a_{end}^{(P)}$. Indeed, for smaller scales, there will be more than one horizon crossing or the crossing will not happen in the slow roll.

Let us then assume that our choice of vacuum state is such that the power suppression is due to physics during the Planck regime and it is hence associated with the scale $k_0$. We will discuss a proposal for a choice of vacuum state with these properties in the next section. In this case, if we assume a power-law behavior for the Hubble parameter during the transition epoch from $a_{end}^{(P)}$ to $a_{inf}$, we easily conclude that 
\begin{equation}\label{k_0}
k_0=k_{inf}\left(\frac{H_P}{H_{inf}}\right)^{\frac{\beta}{\beta-1}} .
\end{equation}
Identifying now $k_{min}$ with $k_0$, instead of $k_{inf}$, it is easy to see that the mismatch of (at least) one order of magnitude in $k_{inf}$ to solve the horizon problem is compensated by the ratio $H_P/H_{inf}$ (we recall that $H_{inf}\approx 0.57 \times 10^{-6} H_P $) if, roughly speaking, $\beta/(\beta-1) \gtrapprox 1/6$. In other words, the problems seem to disappear if $\beta \lessapprox -1/5$. 

To conclude this section, let us determine the analytic solution to the dynamical evolution of the scale factor in the simple model of delayed inflation proposed in Ref. \cite{Liu}, consisting of a de Sitter expansion in the initial Planck regime, followed by a power-law transition period, all before the conventional inflationary epoch with slow-roll behavior. For the Planck period, ending at $\tau_{end}^{(P)}$, the scale factor adopts the standard dependence $a(\tau)= a_{end}^{(P)}/ [1-k_0 (\tau-\tau_{end}^{(P)})]$. In the transition interval, where $H=B a^{\beta-1}$, we first notice that $B= H_P^{\beta} k_0^{1-\beta}$. We separately consider the cases of vanishing and nonvanishing $\beta$. For $\beta=0$, we have
\begin{equation}
a(\tau)=a_{end}^{(P)} e^{k_0 \left[\tau -\tau_{end}^{(P)}\right]}, 
\end{equation}
and, for $\beta\neq0$,
\begin{equation}
a(\tau)= a_{end}^{(P)} \left[1-\beta k_0 (\tau-\tau_{end}^{(P)} )\right]^{-1/\beta }.
\end{equation}
Finally, the scale factor in the conventional inflationary epoch was given in Eq. \eqref{aDesiter}.

In order to find a unique solution from the Planck scale to the end of inflation, we impose continuity of the scale factor at the transition points between the different epochs. In this way, we obtain that, for $\beta=0$,
\begin{equation}
a_{inf}=a_{end}^{(P)} e^{k_0 \left[\tau_{inf} -\tau_{end}^{(P)}\right]} ,
\end{equation}
whereas, for $\beta\neq0$,
\begin{equation}
a_{inf}=a_{end}^{(P)} \left[1-\beta k_0 (\tau_{inf}-\tau_{end}^{(P)} )\right]^{-1/\beta }. 
\end{equation}
Note, on the other hand, that $a_{end}^{(P)}/a_{inf}=(H_P/H_{inf})^{1/(\beta-1)}$ according to our power law, a relation that can be used to derive the value of $\tau_{inf}-\tau_{end}^{(P)}$ from the above formulas. 

\begin{figure}[h!]
\centering
\includegraphics[width=0.9\linewidth]{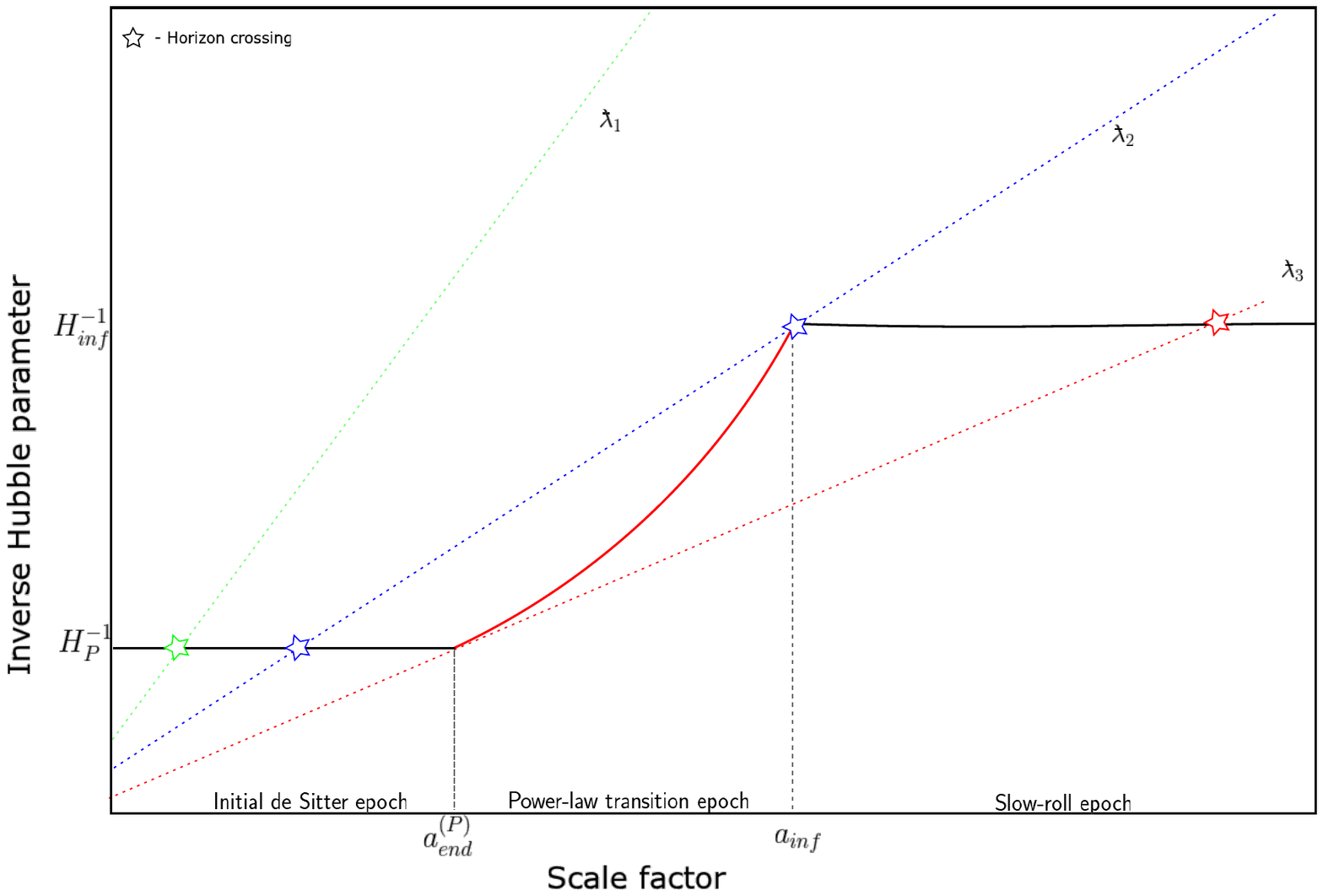}
\caption{Schematic plot illustrating how different modes cross the Hubble horizon. The horizontal solid black lines correspond to the inverse of the (approximately) constant Hubble parameters in the de Sitter evolution during the Planck regime and in the conventional inflationary epoch. The solid red line represents the evolution of the inverse of the Hubble parameter during the transition epoch, for a power law steeper than linear ($\beta < 0 $, with the parametrization used in the text). The plot shows three different angular wavelengths, that exemplify the possible situations. The green dotted line is a mode that exits the horizon during the Planck regime, far away from the conventional slow-roll situation. The dashed blue line represents the first mode that crosses several times the Hubble horizon. For modes with slightly smaller angular wavelengths, the number of crossings becomes equal to three. Finally, the dash-dotted red line corresponds to the supremum of the angular wavelengths that cross the horizon only once and during the slow-roll epoch. Stars represent horizon crossings.}
\label{Horizon crossing}
\end{figure}

In Fig. \ref{Horizon crossing} we present a schematic plot of the considered periods, as proposed in Ref. \cite{Liu}, showing the evolution of the inverse Hubble parameter in terms of the scale factor and illustrating how different perturbation modes cross the horizon a different number of times and in different situations. As we have already commented, we locate the horizon crossing at the point where the physical angular wavelength $\lambdabar= a/k$ equals the Hubble parameter. The curves of constant wavenorm correspond to straight lines in the plot. If we extended the displayed region, all of these curves would eventually intersect at one point, that would ideally mark the origin of the scale factor and the wavelengths. Notice that, for wavelengths beyond $\lambdabar_{P}=1/H_P$, represented as a red line in the figure, the Hubble horizon is either crossed three times or once but during the Planck regime.\\ 

\section{Analytic examples}
\label{Sec4}

Let us present some detailed calculations confirming our arguments, developed in an analytic form for certain examples of physical interest and resting on a concrete proposal for the choice of vacuum, which has been put forward and investigated recently in preinflationary cosmology to try to explain the lack of power at large angular scales \cite{EM,Antonio}. With this aim, we will first find the general solution to the Mukhanov-Sasaki equation \eqref{MSequation} in the three epochs of cosmic evolution introduced in the discussion of delayed inflation. The matching between these three different epochs will be obtained by requiring continuity of the mode solutions up to the first derivative with respect to the conformal time. Finally, we will determine a unique solution in this general family by adhering to the aforementioned proposal for the choice of a vacuum. As our only simplification to solve the perturbations equations analytically, we are going to replace the background-dependent mass of the scalar perturbations in the transition epoch with its expression for tensor perturbations, namely $-a^{\prime\prime}/a$ (in the other two epochs with constant inflaton potential, this replacement is unnecessary). This simplification should be a good approximation at least for power laws in this transition period with $\beta\approx -2$, close to the value corresponding to a kinetically dominated inflaton dynamics \cite{Contaldi,EM}, situation in which the effects of the inflaton potential can be neglected and the scalar and tensor masses coincide in practice\footnote{More generally, the scalar and tensor masses approximately coincide if the so-called Mukhanov-Sasaki potential can be ignored. Its expression in terms of the inflaton potential can be found e.g. in Ref. \cite{mass}.}.   

For the slow-roll inflationary epoch, the general solution has already been provided in Eq. \eqref{solu1}. In the Planck regime with de Sitter expansion, it is also straightforward to obtain the general mode solution. Though mathematically equivalent to the solution for the slow roll, it describes a different epoch with different physics for the perturbations. Using our definition of the associated Fourier scale $k_0$ and calling $\sigma_0=\tau_{end}^{(P)}+1/k_0$, the solution can be expressed as 
\begin{equation}\label{solu2}
v_k=\bar{A}_k \frac{e^{i k(\tau  - \sigma_0)}}{\sqrt{2k}}\left[1 + \frac{i}{k(\tau   - \sigma_0)}\right] + \bar{B}_k\frac{e^{-i k(\tau   - \sigma_0)}}{\sqrt{2k}}\left[1 - \frac{i}{k(\tau  - \sigma_0)}\right] .
\end{equation}

In the transition epoch, we divide our analysis into two cases, corresponding to a linear or a nonlinear behavior of the inverse of the Hubble parameter as a function of the scale factor. Using the expression $-a^{\prime\prime}/a$ for the mass of the perturbations, according to the simplification that we commented above, and the formulas of the previous section for the dynamical evolution of the scale factor, we conclude that the background-dependent mass satisfies  
\begin{equation}\label{sanalyt}
s=- (1+\beta) H^2 a^2=- (1+\beta) k_0^2 \left(\frac{ H_P a}{k_0}\right)^{2\beta}.
\end{equation}
For a linear dependence of the inverse Hubble parameter, corresponding to the constant $\beta=0$, the mass term is $s = - k_0^2$, and the mode equation takes the form of a differential equation for a simple harmonic oscillator with square wavenorm equal to $k^2 - k_0^2$. Consequently, using exponential functions, we obtain the general solution as a linear combination of the form
\begin{equation}
v_k=C_k e^{\sqrt{k_0^2 - k^2} \left(\tau - \tau_{end}^{(P)}\right)} + D_k e^{-\sqrt{k_0^2 - k^2} \left(\tau - \tau_{end}^{(P)}\right)}.
\end{equation}
On the other hand, for $\beta\neq0$, we introduce an auxiliary variable defined as $y= \tau - \tau_{end}^{(P)} -1/(k_0\beta)$. The mass $s$ can then be rewritten as $s=- (1+\beta)y^{-2}/{\beta^2}$, transforming the mode equation into a Bessel differential equation of order $(2+\beta)/(2\beta)$. The general solution of this equation can be expressed as a linear combination of Hankel functions \cite{Abra} of the first and second kind, 
\begin{align}\label{solHa}
v_k=C_k\sqrt{\frac{\pi y}{4}}H^{(1)}_{\frac{2 + \beta}{2\beta}}(ky) + D_k\sqrt{\frac{\pi y}{4}}
H^{(2)}_{\frac{2 + \beta}{2\beta}}(ky) .
\end{align}  

With these expressions and the requirement that the matching of the modes be continuous up to the first derivative in the instants where the model goes from one to another of the considered epochs, we can evaluate the scalar perturbations at the end of slow-roll inflation provided that we give initial conditions for the modes in the initial, Planck epoch. These initial conditions should be those corresponding to a vacuum state, adapted optimally to the background dynamics in the preinflationary and inflationary stages. A criterion to determine such a vacuum in cosmological preinflationary scenarios has been recently proposed in Ref. \cite{BeaDiagonal}. Starting with an asymptotic diagonalization of the Hamiltonian that governs the evolution of the perturbations after a suitable parametrization of them, it is possible to find natural positive solutions in the ultraviolet sector for the perturbations. Employing them, we can construct a state that allows to remove unwanted spurious oscillations in the PPS \cite{BeaPrado}, favoring in this way a low power. Actually, the criterion selects the natural Poincaré vacuum in flat spacetime, and the Bunch-Davies state when the background is (quasi) de Sitter \cite{BeaDiagonal}. Moreover, it also determines a privileged vacuum in kinetically dominated scenarios and in preinflationary epochs with quantum modifications \cite{EM}, and the corresponding PPS presents power suppression with an apparent cutoff related to the characteristic scale of the relevant physics in each case. For all these reasons, we will adopt this criterion in the following to pick out the vacuum of the perturbations. We will call this the NO AHD vacuum, from the initials of non-oscillating (PPS from) asymptotic Hamiltonian diagonalization. 

If the primordial perturbations have quantum origin, as one expects to be the case if they started in the Planck regime, we can determine the corresponding NO AHD vacuum in the following form. The asymptotic diagonalization condition leads in fact to an initial state during the Planck de Sitter expansion which is of Bunch-Davies form, in the sense that it corresponds to a choice of initial data in Eq. \eqref{solu2} such that $|\bar{B}_k|=1$ and $\bar{A}_k=0$ \cite{EM}. The fact that the whole background evolution includes other epochs with different dynamics generically introduces phases that produce rapid oscillations at the end of inflation in the constants $B_k$ and $A_k$ \eqref{solu1} that specify the solution in the slow-roll period. However, it is possible to carry out a Bogoliubov transformation, equivalent to a redefinition of the initial vacuum state, such that the resulting PPS is non-oscillating \cite{EM,Antonio}. The state obtained with this Bogoliubov transformation is the sought NO AHD vacuum.  
   
It is worth commenting that this is not the only proposal existing in the literature for the choice of a vacuum. There are choices of a privileged state for the perturbations that are tailored to specific background behaviors, like in the case of kinetically dominated inflaton evolution \cite{Contaldi}. Other criteria are based on the minimization of quantities related to the renormalized stress-energy tensor (see e.g. Ref. \cite{Handley,Agullo1}). However, the resulting vacuum depends strongly on the instant in which the minimization must occur. It has also been argued that, in the presence of relevant quantum gravity effects, there is only a vacuum state that maintains the Weyl curvature below a certain bound compatible with the uncertainty principle in the whole region with those effects, while respecting a classical behavior at the end of inflation \cite{AG,AG2}. Other proposals introduce a true hard momentum cutoff with an associated vacuum motivated by transplanckian physics or by modified dispersion relations. For example, one can select a state that minimizes the uncertainty in the inflaton configuration field and its momentum at a different instant for each mode, in an inverse relation proportional to the cutoff \cite{Danielsson} (see also the generalization discussed in Ref. \cite{Easther}). Note that this proposal resembles but differs from the commented requirement on the Weyl curvature, which on the other hand is imposed in a whole and the same interval for all modes \cite{AG,AG2}, and not at a single and distinct time for each of them. Another proposal selects the initial state that minimizes a sufficiently adiabatic Hamiltonian on a hypersurface associated with the cutoff \cite{Bozza}. Such Hamiltonian is reached by means of a time-dependent canonical transformation, a procedure which is also employed in the NO AHD criterion. In this last case, the procedure is exploited to the maximum to remove all mode interactions (at least) in the ultraviolet limit, at all times of the interval under consideration. Remarkably, this demand has been proven to eliminate asymptotically the ambiguities on the definition of the inflaton Hamiltonian, and therefore on the choice of vacuum state, in any homogeneous and isotropic (flat) cosmological background \cite{BeaDiagonal}. Finally, as we have mentioned, departures from the Bunch-Davies state can be due to modifications of the dispersion relations. Depending on the dispersion relations and the corresponding initial state, distinct outcomes can be observed in the PPS \cite{Martin,Martin2}.

All these approaches offer valuable tools for discusing corrections to the PPS. Compared to them, a clear distinction of the NO AHD criterion is that it can be formulated on general grounds, for any type of background, and not exclusively for some cosmological scenarios. Using background-dependent canonical transformations, the condition of asymptotic diagonalization can always be made concrete and worked out. This allows one to compare the same physical phenomena in very different situations, adopting in all of them the same criterion for the determination of a vacuum. Furthermore, for standard backgrounds as Minkowski or de Sitter, we recover the natural choice of vacuum state that is conventionally accepted in those cases, as we noted above. In addition, not only the relation of the NO AHD vacuum with adiabatic states is understood, but the former can be defined in circumstances where the adiabatic construction may break down (for instance, for small wavenorms in bouncing cosmologies if the effective mass of the perturbations is approximately given by $-a^{\prime\prime}/a$, since this mass becomes negative at the bounce). We refer the interested reader to Ref. \cite{BeaDiagonal} for technical details related to these properties of the NO AHD vacuum. Lastly, and most remarkably for our discussion, the NO AHD criterion has proven an excellent potentiality to lead to a non-oscillating PPS with a cutoff related to Planck scales without introducing it by hand at any stage of the analysis \cite{EM,Antonio}. It simply arises from an optimal adaptation of the vacuum to the dynamics of the background.

In the rest of this section, we will consider in detail three cases of power-law dependence during the transition epoch that are of especial physical interest, characterized by distinct values of $\beta$.

\subsection{String-dominated transition}

\begin{figure}[h!]
\begin{minipage}[c]{.48\linewidth}
\centering
\includegraphics[width=1\linewidth]{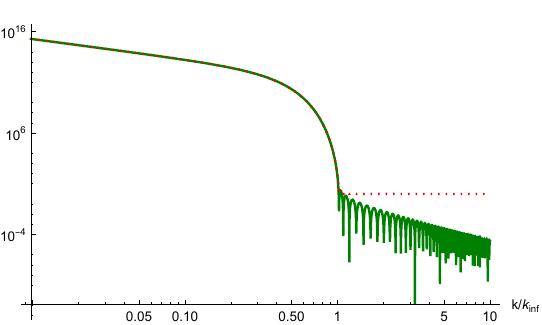}
\caption{Norms $|B_k|$ (dotted red line) and $|A_k|$ (solid green line) of the coefficients \eqref{string} of the mode solution \eqref{solu1} in terms of the rescaled wavenorm $k/k_{inf}$, for the model with string-dominated transition epoch and initial conditions of Bunch-Davies type in the initial Planck epoch. The two axes are in logarithmic scale.}
\label{fig:stringab}
\end{minipage}
\hfill
\begin{minipage}{.48\textwidth}
\centering
\includegraphics[width=1\linewidth]{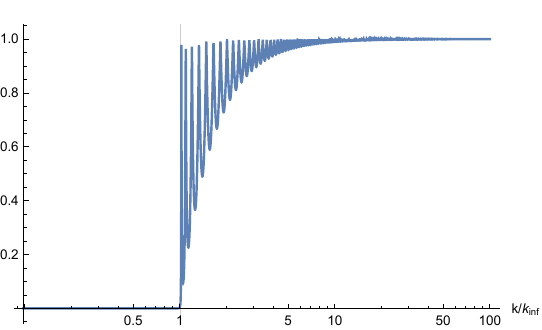}
\caption{PPS of the perturbations determined by the coefficients \eqref{string}, for the model considered in Fig. \ref{fig:stringab}. The two axes are in logarithmic scale. \\ \\ \\}
\label{fig:stringpps}
\end{minipage}
\label{fig:strings}
\end{figure}

A linear dependence of the inverse Hubble parameter on the scale factor is found when the cosmological evolution is dominated by the energy density of a network of strings \cite{Spergel}. In this case, the parameter $\beta$ is zero. Deriving analytic expressions for the perturbations during the slow-roll epoch is straightforward in this situation. As we have commented above, we start in the Planck regime with the initial data $\bar{B}_k=1$ and $\bar{A}_k=0$. A continuous matching up to the first time derivative in the mode solutions leads then to
\begin{align}\label{string}
A_k=i e^{2 i \tilde{k}}\frac{\sinh{\left(\sqrt{1 - \tilde{k}^2} \ln{\tilde{H}_P}\right)}}{2 \tilde{k} \sqrt{1 - \tilde{k}^2} }, \quad & \quad B_k= \cosh{\left(\sqrt{1 - \tilde{k}^2} \ln{\tilde{H}_P}\right)}+i(1 - 2 \tilde{k}^2)\frac{\sinh{\left(\sqrt{1 - \tilde{k}^2} \ln{\tilde{H}_P}\right)}}{2 \tilde{k} \sqrt{1 - \tilde{k}^2} } ,
\end{align}
where we have introduced the rescaled wavenorm $\tilde{k}=k/k_{inf}$ and the rescaled Hubble parameter $\tilde{H}_P=H_P/H_{inf}$. We checked the behaviour of the norm of these coefficients. In the infrared limit $\tilde{k} \rightarrow0$ they both grow unboundedly, with negligible superimposed oscillations. On the other hand, in the ultraviolet limit $\tilde{k}\rightarrow \infty$, the norm of $A_k$ is highly oscillatory and tends to zero, whereas the norm of $B_k$ tends to one. We display the $\tilde{k}$-dependence of the norm of these coefficients in Fig. \ref{fig:stringab}. We also plot the corresponding PPS in Fig. \ref{fig:stringpps}. The fact that the norm of the coefficients (in particular $|A_k|$) oscillates so rapidly for wavenorms $k\geq k_{inf}$ (i.e., $\tilde{k} \geq 1$), indicates that there is no solid motivation for a change of vacuum by means of a Bogoliubov transformation leading from $B_k$ and $A_k$ to new coefficients equal to their norms. This transformation would not only remove spurious power contributions coming from large oscillations in the phases of the mode solutions, but also relevant information about the background captured in the variation of the norms, as discussed in Ref. \cite{EM}. This is why we do not perform in this case a change to a vacuum with a non-oscillating spectrum. For our discussion, the change is not necessary in fact. We already see in Fig. \ref{fig:stringpps} that the PPS shows a cutoff, but this cutoff appears at the scale corresponding to the beginning of slow-roll inflation, namely $k=k_{inf}$, and therefore cannot resolve the problem that we discussed in Sec. \ref{Sec3}. This result agrees with our previous arguments in that section, that indicated the need that $\beta \lessapprox - 1/5$. Notice also that the scale invariance of the PPS is regained at large wavenorms, where it coincides with the PPS of a Bunch-Davies state in the slow roll (the normalized PPS becomes equal to one in this sector).

\subsection{Radiation-dominated transition}

If radiation dominates the transition from the Planck regime to the slow roll, the expansion rate (proportional to the square of the energy density according to Friedmann equation) follows the scaling behavior  $H\propto a^{-2}$. Hence, this case corresponds to a power-law transition with $\beta=-1$. It is again straightforward to use our analytic expressions to derive the value of the constant coefficients $B_k$ and $A_k$ that determine the mode solution at the end of inflation, starting with initial data at the Planck regime given by $\bar{B}_k=1$ and $\bar{A}_k=0$. Employing the rescaled quantities $\tilde{k}$ and $\tilde{H}_P$, and recalling the definition of $k_0$ given in Eq. \eqref{k_0} (particularized to $\beta=-1$), we obtain  
\begin{align}
A_k &= \frac{1}{4 \tilde{k}^4 } \left\{e^{\frac{2 i \tilde{k}}{\sqrt{\tilde{H}_P}}} \left[\tilde{H}_P-2 i \tilde{k}  \left(\sqrt{\tilde{H}_P}-i \tilde{k}\right)\right]+ e^{2 i \tilde{k}} \tilde{H}_P  \left[-1+2 \tilde{k} (i+\tilde{k})\right] \right\} , \nonumber  \\
B_k &= \frac{1}{4 \tilde{k}^4 } \left\{e^{-2 i \tilde{k} \left(1-\frac{1}{\sqrt{\tilde{H}_P}}\right)}  \left[-1+2 \tilde{k} (-i+\tilde{k})\right] \left(- \tilde{H}_P+ 2i \tilde{k}
\sqrt{\tilde{H}_P} + 2 \tilde{k}^2 \right) - \tilde{H}_P  \right\} . \label{radiat}
\end{align}

\begin{figure}[h!]
\begin{minipage}{.48\textwidth}
\centering
\includegraphics[width=1\linewidth]{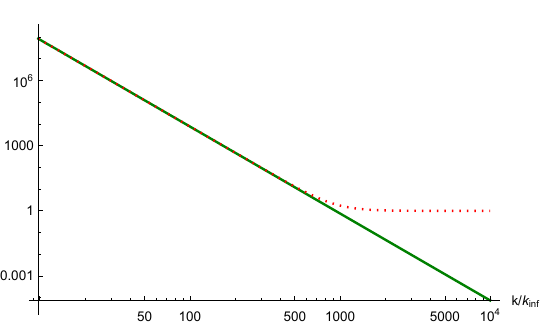}
\caption{Norms $|B_k|$ (dotted red line) and $|A_k|$ (solid green line) of the coefficients \eqref{radiat} of the mode solution \eqref{solu1} in terms of the rescaled wavenorm $k/k_{inf}$, for the model with radiation-dominated transition epoch and initial conditions of Bunch-Davies type in the initial Planck epoch. The two axes are in logarithmic scale.}
\label{fig:ard}
\end{minipage}
\hfill
\begin{minipage}{.48\textwidth}
\centering
\includegraphics[width=1\linewidth]{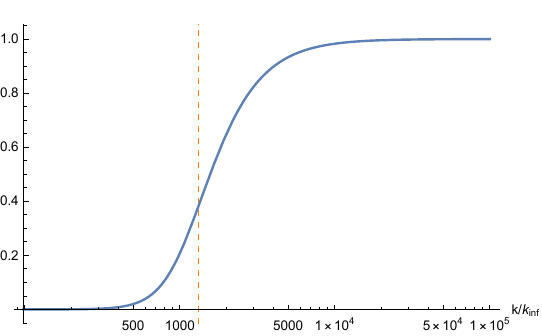}
\caption{Non-oscillating PPS of the perturbations obtained with a Bogoliubov transformation from the coefficients \eqref{radiat}, for the model considered in Fig. \ref{fig:ard}. The two axes are in logarithmic scale. The dashed orange line marks the location of $k_0$. \\ \\}
\label{fig:ppsard}
\end{minipage}
\label{fig:rd}
\end{figure}

In Fig. \ref{fig:ard} we show the norms of these coefficients as functions of $k/k_{inf}$. The norms vary smoothly in all the considered range of wavenorms, without rapid oscillations. They grow unboundedly in the infrared, whereas in the ultraviolet limit $\tilde{k}\rightarrow \infty$ the norm of $A_k$ tends to zero, and consequently the norm of $B_k$ tends to the unit. 

Following our prescription for the choice of vacuum, we then implement a Bogoliubov transformation that removes the phase oscillations and leads to a non-oscillating PPS. We display this PPS in Fig. \ref{fig:ppsard}. Notably, the spectrum presents power suppression in a cutoff that corresponds to a wavenorm much greater than the one associated to the beginning of the slow roll, $k_{inf}$. More precisely, the cutoff wavenorm is three orders of magnitude larger. This result is in full agreement with our estimation of the cutoff in Eq. \eqref{k_0}. Given relation \eqref{Hintvalue} and that $\beta=-1$, we get the estimation $k_0 \approx 1.3 \times 10^{3} k_{inf}$, remarkably similar to the value obtained with our choice of vacuum and its non-oscillating PPS.  

\subsection{Kinetically dominated transition}

Finally, we study the case of a transition epoch in which the energy density of the inflaton is dominated by the kinetic contribution. This scenario has attracted a lot of attention in the search for alternatives to standard inflationary models (see e.g. Refs. \cite{Contaldi,EM} and references therein), and it has been suggested that an intermediate epoch with this behavior may be compatible with observational data, at least in some quantum gravity inspired cosmologies \cite{tango,Antonio}. From our discussion in Sec. \ref{Sec2}, it is not difficult to see that this case corresponds to a power-law behavior of the Hubble parameter with $\beta=-2$. We follow the same line of arguments as in the two cases considered above, starting with initial conditions of Bunch-Davies type in the Planck epoch, namely $\bar{B}_k=1$ and $\bar{A}_k=0$. In this way, we arrive at a mode solution \eqref{solu1} at the end of inflation determined by the coefficients
\begin{align}
A_k&=\frac{\pi e^{i \tilde{k} \left(1+\tilde{H}_P^{-2/3}\right)}} {16 \tilde{k} \tilde{H}_P^{1/3}} \left[\tilde{k} H_0^{(2)}\Big(\frac{\tilde{k}}{2}\Big)+i (\tilde{k}+i) H_1^{(2)} \Big(\frac{\tilde{k}}{2}\Big)\right] \left[ \tilde{k}H_0^{(1)}\left( \frac{\tilde{k} \tilde{H}_P^{-2/3}}{2} \right) - \left(\tilde{H}_P^{2/3}-i \tilde{k}\right) H_1^{(1)}\left( \frac{\tilde{k} \tilde{H}_P^{-2/3}}{2} \right) \right] \nonumber \\
 &  - (1) \leftrightarrow (2)   , \nonumber \\
B_k=&\frac{\pi  e^{-i \tilde{k} \left(1-\tilde{H}_P^{-2/3}\right)} }{16 \tilde{k} \tilde{H}_P^{1/3}} \left[ \tilde{k} H_0^{(2)}\Big( \frac{\tilde{k}}{2} \Big)-i (\tilde{k}-i)
H_1^{(2)}\Big(\frac{\tilde{k}}{2}\Big) \right] \left[ \tilde{k}  H_0^{(1)}\left( \frac{\tilde{k} \tilde{H}_P^{-2/3}}{2} \right) - \left( \tilde{H}_P^{2/3}-i \tilde{k} \right) H_1^{(1)}\left(\frac{\tilde{k} \tilde{H}_P^{-2/3}}{2 }\right) \right]  \nonumber \\
 &  - (1) \leftrightarrow (2)   .
\label{kinetic}
\end{align}  
We have used the notation $(1) \leftrightarrow (2)$ to denote terms obtained from those displayed explicitly by interchanging Hankel functions of the first and second kind.

\begin{figure}[h!]
\begin{minipage}{.48\textwidth}
\centering
\includegraphics[width=1\linewidth]{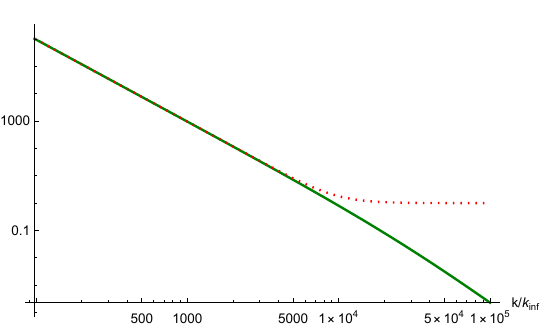 }
\caption{Norms $|B_k|$ (dotted red line) and $|A_k|$ (solid green line) of the coefficients \eqref{kinetic} of the mode solution \eqref{solu1} in terms of the rescaled wavenorm $k/k_{inf}$, for the model with kinetically dominated transition epoch and initial conditions of Bunch-Davies type in the initial Planck epoch. The two axes are in logarithmic scale.}
\label{fig:akd}
\end{minipage}
\hfill
\begin{minipage}{.48\textwidth}
\centering
\includegraphics[width=1\linewidth]{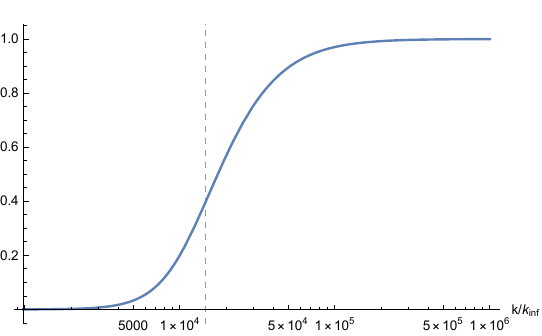}
\caption{Non-oscillating PPS of the perturbations obtained with a Bogoliubov transformation from the coefficients \eqref{kinetic}, for the model considered in Fig. \ref{fig:akd}. The two axes are in logarithmic scale. The dashed orange line marks the location of $k_0$. \\ \vspace{.2cm}}
\label{fig:ppskd}
\end{minipage}
\label{fig:kd}
\end{figure}

We display the wavenorm dependence of the norms of these coefficients in Fig. \ref{fig:akd}. We see that they have a smooth dependence without rapid oscillatory variations. We then compute the associated non-oscillating PPS, obtained by means of a Bogoliubov transformation in our vacuum state which eliminates unwanted oscillatory phases. We plot this PPS in Fig. \ref{fig:ppskd}. We note the clear power suppression, which occurs at a wavenorm scale that is significantly larger than $k_{inf}$. In the present case, the cutoff scale is approximately four orders of magnitude larger. Again, this is in remarkable coincidence with the estimation provided by Eq. \eqref{k_0}, with $\beta=-2$, and relation \eqref{Hintvalue}, which give $k_0 \approx 1.4 \times 10^{4} k_{inf}$. This suggests that power suppression occurs at wavenorm scales related (by means of the cosmological background evolution) with the Planck scale, and not with the scale of the beginning of inflation. The difference is indeed enough to resolve the challenges that the observational CMB data imply for the inflationary mechanism as the origin of the temperature anisotropies.

\section{Discussion}
\label{Sec6}

Observational data of the CMB determine a scale for the beginning of standard slow-roll inflation that differs considerably from the natural Planck scale of chaotic inflation or other theoretical frameworks inspired by quantum considerations. This discrepancy of scales poses serious theoretical problems even in delayed inflationary models which propose transitions between periods of expansion at those two different scales. Actually, such models introduce a (nearly hard) cutoff in the PPS compatible with present observations. However, if this cutoff is assigned to the perturbation mode that crossed the Hubble horizon just at the beginning of slow-roll inflation, it has been argued that there is a mismatch of at least one order of magnitude to reach a satisfactory explanation of the averaged isotropy that we observe in the CMB and the related homogeneity supported by it \cite{Liu}. 

In this article, we have pointed out that this problem disappears if one allows for a vacuum of the perturbations different from the standard Bunch-Davies state at the slow roll. The conflict arises from the identification of the cutoff scale and the scale at the beginning of the slow roll, a sustainable identification only if the perturbations are in a Bunch-Davies state when they cross the horizon during the entire slow-roll epoch. However, a preinflationary era can generate excited modes that already crossed the horizon before that epoch, leading to departures from the Bunch-Davies behavior. 

In a simple situation in which an initial de Sitter expansion in the Planck regime is connected with the standard slow-roll epoch by a transition period, with a Hubble parameter that displays a power-law dependence on the scale factor of the form $H\propto a^{\beta-1}$, a straightforward calculation confirms that modes with wavenorms larger than the one associated to the beginning of the slow roll can be excited with respect to the Bunch-Davies state. Therefore, a suitable choice of vacuum for such modes that abandons the Bunch-Davies criterion can produce power suppression at scales which restore the consistency with the expansion needed for homogeneity. We have computed this difference of scales and demonstrated that it is enough to solve the commented problem if the parameter $\beta$ of the transition period is (approximately) smaller than $-1/5$. Moreover, we have carried out an analytic study of the PPS based on a specific criterion to determine the vacuum of the perturbations. This criterion selects the vacuum by means of an asymptotic diagonalization of the Hamiltonian that generates the Heisenberg dynamics of the perturbations \cite{BeaDiagonal} and removes spurious, rapidly varying oscillations in the PPS. The only simplification that we have made in this analysis is to replace the background-dependent mass of the scalar perturbations by the mass corresponding to tensor perturbations, which is much more manageable. The important fact is that it is then possible to prove analytically that the perturbations at the end of inflation display a clear cutoff in the PPS, and that this cutoff approximately coincides with the value derived with our general arguments. We have studied in detail the case of a power law corresponding to cosmic strings, radiation, or the kinetic contribution of the inflaton. In the first case, the discrepancy between the cutoff and the scale at the beginning of the slow roll is not sufficient to solve the apparent conflict of scales, but in the two other cases, the problem disappears. Moreover, in the case of a kinetically dominated expansion, there is no simplification in our replacement of the mass of the scalar perturbations with the tensor one, because both masses coincide.

Our formula (3.5), together with the values of the Hubble parameter at the slow roll and the cutoff extracted from the observational data, allows us to estimate quantitatively the difference between the two aforementioned scales, namely, the scale $k_{inf}$ at the beginning of standard inflation and the scale $k_{min}$ of the cutoff. Identifying the latter with $k_0$, i.e. the infimum scale for which there is no horizon crossing before slow-roll inflation, we get $k_{inf}=k_{min}(H_{inf}/H_P)^{\beta/(\beta-1)}$. Recalling that $H_{inf}/H_P\leq 1$, it is easy to check that this quantity decreases when $\beta$ becomes more negative. For the two interesting cases of a transition period dominated by radiation ($\beta=-1$) or by the kinetic energy of the inflaton ($\beta=-2$), we respectively obtain, in terms of the pivot scale $k_{\star}$,
\begin{align}\label{scalefactors}
k_{inf} \approx 4.8 \times 10^{-6} k_{\star} \quad {\rm (radiation),} \qquad & \qquad k_{inf} \approx 4.3 \times 10^{-7} k_{\star} \quad {\rm (kinetic)}.
\end{align} 
Clearly, these values are below the observational bound $k_{inf}\leq 1.45 \times 10^{-3} k_{\star}$, resolving the problem as we said. 

Our analysis has been developed in the framework of general relativity. To conclude our discussion, it is worth commenting on a more general scenario with quantum gravity corrections. In LQC, as we have already mentioned, the universe bounces in the Planck regime avoiding the appearance of a Big Bang \cite{APS}. At this bounce, the Hubble parameter vanishes effectively. The bounce is followed by a Planck epoch of superinflation, in which the Hubble parameter grows fast to Planck values \cite{AS}. After this Planck era, the typical backgrounds explored in LQC enter a period of kinetically dominated expansion in which the classical description of general relativity holds. This kinetic period is followed by a standard inflationary period, assuming the presence of an inflaton in the cosmological model. As we see, the early universe follows an evolution that is very similar to that found in the simple situation with three epochs analyzed in this article. Apart from restricting the power-law transition to an epoch of kinetically dominated dynamics, with the specific value $\beta=-2$, the only difference is that the initial epoch of de Sitter expansion in the Planck regime is substituted by a more complicated expansion. But the rest of the features are notably similar. Moreover, the PPS for LQC has been studied in detail choosing the NO AHD vacuum \cite{BeaDiagonal,EM,Antonio}. It has been shown that the spectrum of the perturbations displays a cutoff at a scale in direct correspondence with the Planck scale of superinflation. The spectrum has also been studied with another concrete proposal for the determination of the vacuum state which seems to lead also to some power suppression at a similar scale \cite{tango}, namely the proposal of Refs. \cite{AG,AG2}. The fact that an effective cutoff appears at a scale associated to the Planck epoch (especially for the NO AHD vacuum) allows us to regard this scenario with quantum geometry effects as an extension of the relativistic case contemplated in this work.    

\acknowledgments

The authors are grateful to B. Elizaga Navascu\'es and T. Paw{\l}owski for discussions. This work was supported by the Polish National Center for Science (Narodowe Centrum Nauki -- NCN) under grant OPUS 2020/37/B/ST2/03604 and by the Spanish grant PID2020-118159GB-C41 funded by MCIN/AEI/10.13039/501100011033/.

\end{document}